\documentclass[reprint, showpacs, preprintnumbers, amsmath, amssymb, aps, prb,url,]{revtex4-1}
\usepackage{color}
\usepackage{graphicx}
\usepackage{bm}
\usepackage{dcolumn}
\usepackage{sidecap}
\usepackage{amsmath}
\sidecaptionvpos{figure}{m}

\begin{document}

\bibliographystyle{apsrev4-1}

\title{Hybrid excitations due to crystal-field, spin-orbit coupling and spin-waves in LiFePO$_4$}

\author{Yuen Yiu$^{a}$, Manh Duc Le$^b$, Rasmus Toft-Peterson$^c$, Georg Ehlers$^d$, Robert McQueeney$^{a}$, and David Vaknin$^{a}$}
\affiliation{
$^a$DMSE, Ames Laboratory and Department of Physics and Astronomy, Iowa State University, USA\\
$^b$ISIS Neutron and Muon Source, Rutherford Appleton Laboratory, UK\\
$^c$Helmholtz-Zentrum Berlin f\"ur Materialen und Energie, Germany\\
$^d$Quantum Condensed Matter Division, Oak Ridge National Laboratory, USA\\
}

\date{\today}

\begin{abstract}

We report on the spin waves and crystal field excitations in single crystal LiFePO$_4$ by inelastic neutron scattering over a wide range of temperatures, below and above the antiferromagnetic transition of this system.  In particular, we find extra excitations below $T_N=50$ K that are nearly dispersionless and are most intense around magnetic zone centers. We show that these excitations correspond to transitions between thermally occupied excited states of Fe$^{2+}$ due to splitting of the $S=2$ levels that arise from  crystal field and spin-orbit interaction. These excitations are further amplified by the highly distorted nature of the oxygen octahedron surrounding the iron atoms.  Above $T_N$,  magnetic fluctuations are observed up to at least 720~K, with additional excitation around 4 meV, likely caused by single-ion splittings through spin-orbit and crystal field interactions. The latter weakens slightly at 720~K compared to 100~K, which is consistent with calculated cross-sections using a single-ion model.  Our theoretical analysis, using the MF-RPA model, provides both detailed spectra of the Fe $d-$ shell and estimates of the average ordered magnetic moment and $T_N$. By applying the MF-RPA model to a number of existing spin-wave results from other Li$M$PO$_4$ ($M=$ Mn, Co, and Ni), we are able to obtain reasonable predictions for the moment sizes and transition temperatures.
\end{abstract}
\maketitle

\section{Introduction}


Various members of the lithium-orthophosphates have gained renewed attention in the last decade as candidates of electrodes in lithium based rechargeable batteries \cite{park2010,padhi1997,chung2002,thomas2003} by virtue of their high Li-ion conductivity through channels that are present in their olivine crystal structure\cite{park2010,yang2011,nishimura2008}. Thus, although most research efforts have recently been focused on their electro-chemical properties, they have long been known to exhibit intriguing magnetic properties.  In particular, the  transition-metal based lithium-orthophosphates display strong magnetoelectric (ME) effect, where an applied magnetic field induces an electric polarization and vice-versa, applied electric field induces magnetization. Naturally, the ME effect is invoked by the coupling among orbital, magnetic, and electrostatic degrees of freedom, that have been at the forefront of recent research in condensed matter physics. Prominent examples of such coupling have been found in the iron- and copper-based unconventional superconductors\cite{dai2012, johnston2010}, the giant magnetoresistance in Mn-based oxides\cite{pickett1996,ramirez1997}, or the magnetoelectric effect in transition metal oxides\cite{nan2008,li2008}. Often, interrelated magnetic, electric and/or structural transformations raise the question of their origin, but it is generally accepted that the spin-orbit coupling (SOC) drives a secondary transition, namely magnetic or structural that follows structural or magnetic respectively. The SOC can also lead to higher order asymmetric exchange terms such as the Dzyaloshinskii-Moriya interaction and can lift the degeneracy of crystal field levels. In the case of the lithium-orthophosphates, it has been observed that the relative strength of the ME effect correlates with the effective total orbital moment with LiCoPO$_4$ ($L=3$) and LiMnPO$_4$ ($L=0$) displaying the largest and smallest ME coefficients \cite{rivera1994a,Rivera1994b,vaknin2002, mercierthesis}. Here, we report on the magnetic excitations of LiFePO$_4$ and expand on recent elastic and inelastic neutron studies of the lithium-orthophosphates\cite{jensen2009,toftpetersen2011,toftpetersen2012}.
Similar to other lithium-orthophosphates, LiFePO$_4$ possesses an intricate magnetic structure, where the Fe$^{2+}$  moments ($S=2; L=2$) order antiferromagnetically at $T_N =  50$ K with moments pointing mainly along the $b$-direction\cite{liang2008,santoro1967}. However, subsequent studies have revealed zero field spin canting along $a$ and $c$, which are both forbidden by \textit{Pnma} symmetry, hinting that the crystal structure symmetry might be lower than \textit{Pnma} below $T_N$\cite{toftpetersen2015}. The observed spin-canting implies the presence of Dzyaloshinsky-Moriya (DM) interactions, which can be linked to the ME response. The local symmetry of the magnetically ordered Fe sublattice may also be reflected by the crystal field level structure of Fe(2+)\cite{xiao2013}.

Several inelastic neutron scattering (INS) efforts have measured the spin wave spectrum in LiFePO$_4$ in restricted regions of the Brillouin zone\cite{toftpetersen2015,li2006}. The most comprehensive model for LiFePO$_4$ deduced from these measurements includes 5 exchange interaction terms (1 in-plane nearest neighbour, 2 in-plane next-nearest-neighbours, and 2 out-of-plane interactions), and 2 single-ion anisotropy terms (along $a$ and $c$). In this study we complete the INS picture for spin waves measured along all directions $a$, $b$, and $c$ and compare our results with the existing Hamiltonian. We also report on new low-energy excitations found below the spin waves excitations, which persist from below $T_N$  up to 720~K. We argue that such excitations are due to single-ion splitting of the $S=2$ manifold from the crystal field, spin orbit and ordered moment exchange field. Modifications of the existing spin Hamiltonian model include these hybrid interactions to account for the new excitations.

\section{Experimental details}
\begin{figure*}
  \centering
 \includegraphics[width = 0.8\textwidth]{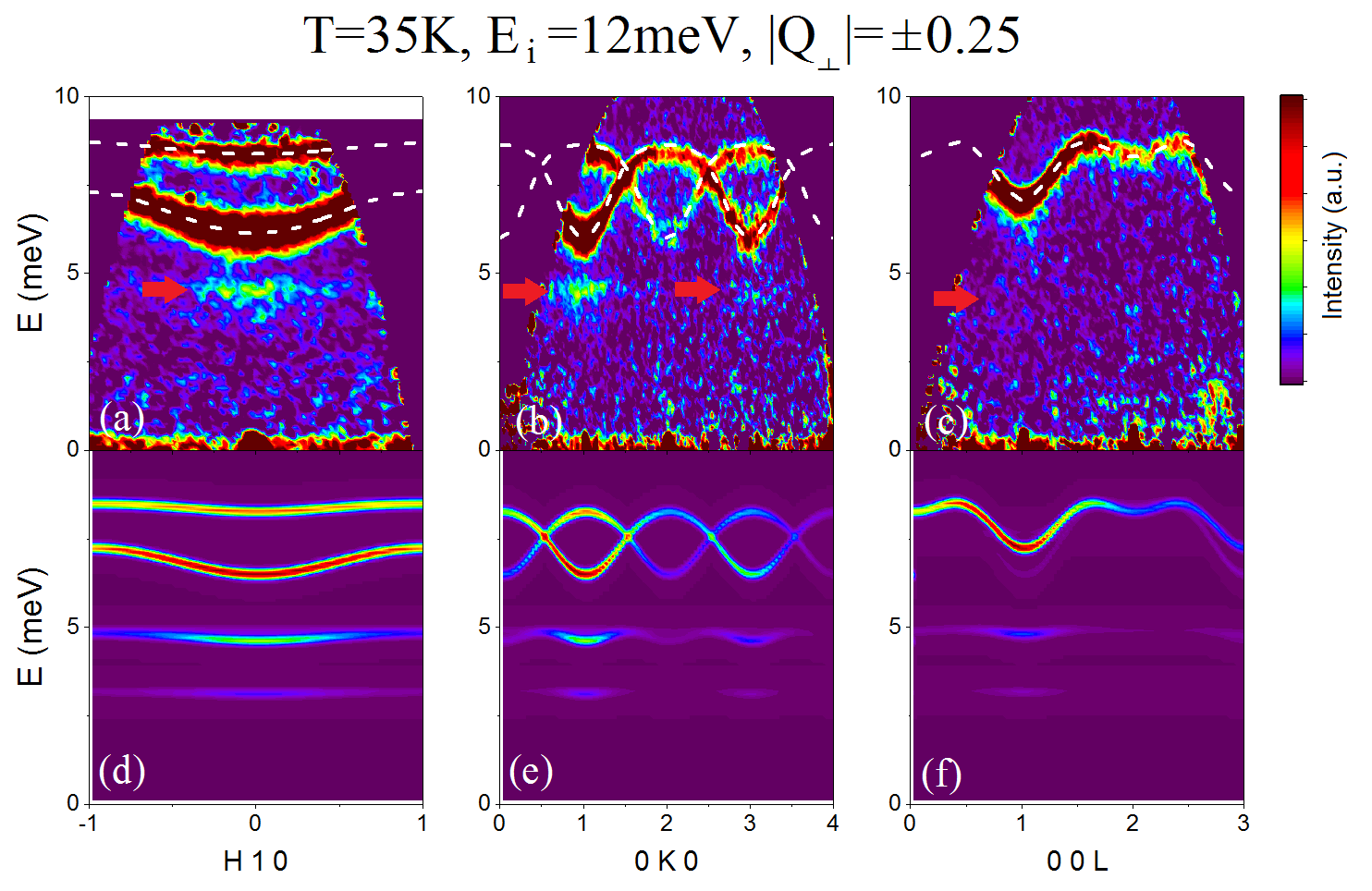}
      \caption{\label{calculatedspinwave}(color online) (a-c): Contour plots of energy transfer from the inelastic neutron scattering (INS) data collected at $T  =  35$ K along ($H10$), ($0K0$), and ($00L$). The dashed  lines are calculated linear spin wave dispersions based on the model described in the text. The red arrows indicate extra excitations visible around 4.5 meV that corresponds to Zeeman splitting levels by internal mean field induced by the ordered moments. (d-f): Virtual INS data calculated based on the same model using mean-field random-phase approximation. As can be seen the simulation correctly predicts the {\it hybrid} extra excitations near 4.5 meV in addition to an extra less intense excitation near 3 meV, both originating from {\it internal}-Zeeman splitting.}
\end{figure*}

\begin{figure}
\includegraphics[width = 82mm]{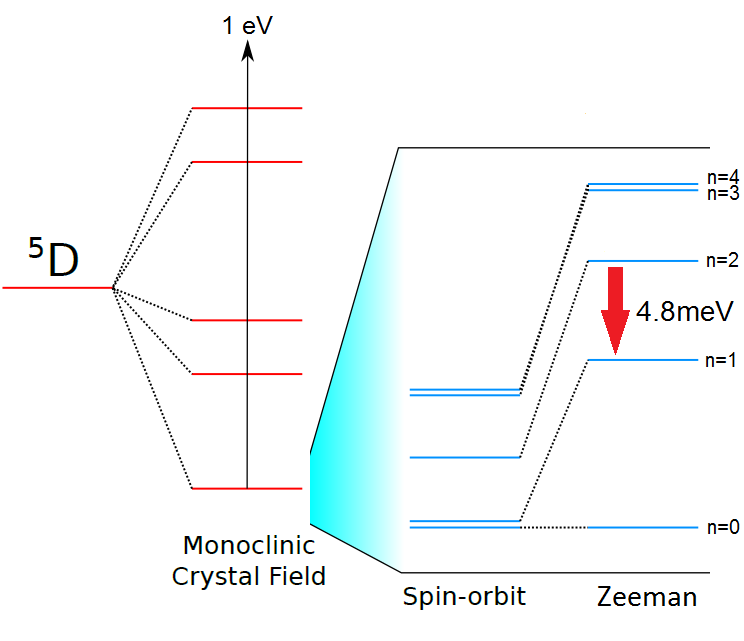}
  \caption{\label{crystal_fields}(color online) Multi-electron states arising from the crystal field splitting, each of these orbital singlets has a five-fold spin degeneracy. The ground state multiplet is further split as indicated by the blue lines, either by spin-orbit coupling or internal magnetic field when the system is magnetically ordered. }
\end{figure}
For the inelastic neutron scattering experiment, a single crystalline sample of LiFePO$_4$ was used.  The crystal was selected from a batch of single crystals synthesized using the standard flux growth method, using the same recipe prescribed by Ref.~\onlinecite{toftpetersen2015}. The high quality single crystal weighs approximately 200mg, and its structure and stoichiometry were confirmed by laboratory X-ray and by single crystal neutron diffraction\cite{toftpetersen2015}. Inelastic neutron scattering (INS) data were collected at the Cold Neutron Chopper Spectrometer (CNCS)\cite{ehlers2011} of the Spallation Neutron Source, Oak Ridge National Laboratory. The sample was aligned with the $bc$-plane horizontal with some detector coverage along the $a$ (vertical) direction. The incident neutron energy was set at E$_i$  =  12 meV for the optimal resolution and flux for the $(\mathbf{Q},\omega)$ region of relevant interest. INS data was collected at three separate temperatures  T =  35, 100, and 720~K. 
\section{Results and discussion}

Figure.\ \ref{calculatedspinwave}(a-c) shows contour plots of the INS data collected at $T  =  35$~K along the $(H10)$, $(0K0)$, and $(00L)$ reciprocal space directions, with an incident neutron energy of $E_i$  =  12 meV and integrated over a range of $|Q_\perp|$ of $\pm 0.25$ reciprocal lattice units. Along $(0K0)$ two spin-wave branches are clearly visible, as expected by the known antiferromagnetic structure, which contains spins precessing perpendicular to the moment direction (along $b$)\cite{toftpetersen2015}. One branch where the two oppositely aligned spins precess in-phase, and one branch where the two precess out-of-phase.  Also shown in Figure~\ \ref{calculatedspinwave} are dotted lines obtained from linear spin wave calculations using the existing spin Hamiltonian described below\cite{toftpetersen2015}. 
\begin{equation}
\label{eq1}
\mathcal{H}  = \frac{1}{2} \sum_{i,j} J_{ij} \textbf{S}_i\cdot \textbf{S}_j + \mathcal{H}_{\mathrm{SIA}}
\end{equation}
The Hamiltonian considers 5 exchange interaction terms, among them are one nearest neighbor $J_{bc}$, two next nearest neighbors $J_{b}$ and $J_{c}$, and two out-of-plane interactions $J_{ab}$ and $J_{ac}$. Previous reports based on single crystal data have found magnetic anisotropy along $a$ and $c$\cite{liang2008,li2006}, and as a result the Hamiltonian requires the inclusion of single-ion anisotropy (SIA) terms $D_a$ and $D_c$, allowing for an anisotropic hard plane while the easy axis $b$ direction is set by $D_b$ = 0. For the linear spin wave calculation shown here, the $J$ and $D$ values from previously published inelastic single crystal data of the same crystal are used\cite{toftpetersen2015}. These values are listed in Table~\ref{spintable}. 

\begin{table}[htb]
\begin{center}
\caption{\label{spintable} Exchange interactions $J_{ij}$ and single ion anisotropy $D_a$ and $D_c$ (in meV) used in calculating the spin-waves shown in Figure~\ \ref{calculatedspinwave}. $H_{\mathrm{mf}}$ is calculated from the $J_{ij}$'s in the mean-field approximation } 
\begin{tabular}{| r | c | c |}
    \hline
             &  Fig.\ref{calculatedspinwave}(a-c)  &  Fig.\ \ref{calculatedspinwave}(d-f) \\
             &   Ref.\ \onlinecite{toftpetersen2015} &  This study \\
	\hline
    $J_{bc}$          &  0.77 (2) &  0.46 (2) \\
    $J_{b}$           &  0.30 (2) &  0.09 (1) \\
    $J_{c}$           &  0.14 (2) &  0.01 (1) \\
    $J_{ab}$          &  0.14 (2) &  0.09 (1) \\
    $J_{ac}$          &  0.05 (1) &  0.01 (1) \\
    $D_{a}$           &  0.62 (1) &  0.86 (2) \\
    $D_{c}$           &  1.56 (4) &  2.23 (2) \\
    $H_{\mathrm{mf}}$ &  4.84     &  3.69     \\
    \hline
\end{tabular}
\end{center}
\end{table}

\subsection{Zeeman splitting}

As illustrated in Figure~\ref{calculatedspinwave}(a-c), the linear spin wave calculations (white dotted lines) track the measured magnon dispersion closely, however it does not account for the extra excitations visible near 4.5 meV indicated by the red arrows. These extra excitations are nearly dispersionless in energy and centered around the magnetic zone centers. We argue that these extra excitations arise from transitions from a thermally populated excited state of the spin $S=2$ multiplet that are not considered by spin wave theory, which is a $T=0$ theory and maps the spin ground state to a vacuum state and transitions from (to) this state to (from) the $n^{\mathrm{th}}$ excited state to creation (annihilation) of $n$ magnons. The  4.5 meV excitation corresponds to a transition between two excited states, $n=1$ and $n=2$, as illustrated schematically in Figure~\ref{crystal_fields}. We thus use the mean-field random-phase approximation (MF-RPA)\cite{jensen1991} to calculate the higher temperature magnetic spectrum.

Figure~\ref{crystal_fields} shows schematically how the different single-ion interactions splits the $d$-electron energy levels. The largest energy splitting, of the order of an electron-volt, is from the crystal field, which only acts on the orbital angular momentum, leaving the spin states degenerate. In the lithium orthophosphates, the Fe$^{2+}$ ions are located at a low (monoclinic $C_s$) symmetry site, so the crystal field splitting lifts all orbital degeneracy resulting in five orbital singlets, which each have a five-fold spin degeneracy. The on-site spin-orbit interaction is much weaker than the crystal field interaction in transition metals and only splits the spin states by a few meV. Finally, in the ordered phase, the ordered moments generate a internal magnetic field which further Zeeman splits the spin energy levels. 

The effects of the spin-orbit and crystal field interactions on the $S=2$ levels can be parametrized by the single-ion anisotropy parameters $D_a$ and $D_c$, whilst the Heisenberg term leads to a Zeeman-like interaction, in the mean-field (MF) approximation. Thus we can rewrite equation~\ref{eq1} in the MF-approximation as an effective single-ion Hamiltonian:
\begin{equation}
\label{eq2a}
\mathcal{H} ^{(1)}_{\mathrm{mf}} = \mathcal{H}_{\mathrm{SIA}} + \mathcal{H}_{\mathrm{Zeeman}}
\end{equation}

\noindent where, after choosing the moment direction as the quantization axis, that is $z||b$,

\begin{eqnarray}
\mathcal{H}_{\mathrm{SIA}} &=& D_a \hat{S}_{x}^2 + D_c \hat{S}_{y}^2 
\\
\mathcal{H}_{\mathrm{Zeeman}} &=& - H_{\mathrm{mf}} \hat{S}_{z}
\end{eqnarray}

\noindent where $D_a$, $D_c$ and $H_{\mathrm{mf}}$ (which depends on the $J_{ij}$ parameters) are given in Table~\ref{spintable}. Diagonalising this Hamiltonian in the $|S_z\rangle$ basis with the Zeeman term results in the level scheme shown in Table~\ref{zeemanlevels} for the ordered (AFM) phase, whereas setting $H_{\mathrm{mf}}=0$ gives the level scheme shown for the paramagnetic phase. Furthermore, inspection of the results of diagonalising the Hamiltonian with different values of $D_a$, $D_c$ and $H_{\mathrm{mf}}$ shows that when $H_{\mathrm{mf}}>(D_a+D_c)$, the difference in energy between the $n=1$ and $n=2$ levels is $\Delta_{12}=H_{\mathrm{mf}}+(D_a+D_c)/2$.

\begin{table}[htb]
\begin{center}
    \caption{\label{zeemanlevels} Calculated level scheme due to the crystal field and spin-orbit interactions, parameterized by the single-ion parameters $D_a=0.86$, $D_c=2.23$ meV, in the paramagnetic phase and in the magnetically ordered phase with an additional Zeeman splitting term parameterised by $H_{\mathrm{mf}}=3.69$ meV.} 
\begin{tabular}{| r | c | l |}
\hline 
       &Energy & \\ 
       &(meV)& \ \ \ \ \ \ $|\ \ S_z\rangle$ components \\ 
\hline
\multicolumn{3}{| c |}{Paramagnetic Phase} \\
\hline
n=0 & 0     & $\phantom + 0.67\ (|-2\rangle + |+2\rangle)$ + 0.32\ $|0\rangle$ \\
1   & 0.81  & \ $1/\sqrt{2}\ (|-2\rangle - |+2\rangle)$       \\
2   & 3.39  & \ $1/\sqrt{2}\ (|-1\rangle + |+1\rangle)$       \\
3   & 7.49  & \ $1/\sqrt{2}\ (|-1\rangle - |+1\rangle)$       \\
4   & 7.79  &     $-0.23\ (|-2\rangle+|+2\rangle)+0.95\ |0\rangle$            \\
\hline
\multicolumn{3}{| c |}{Ordered Phase} \\
\hline
n=0 & 0     & 0.99\ $|+2\rangle$               + 0.12\ $|\phantom+ \   0\rangle$ \\
1   & 8.00  & 0.97\ $|+1\rangle$               + 0.25\ $|-1\rangle$\\
2   & 12.73 & 0.80\ $|\phantom+ \ \,0\rangle$  + 0.60\ $|-2\rangle$\\
3   & 16.21 & 0.80\ $|-2\rangle$               + 0.60\ $|\phantom+ \    0\rangle$\\
4   & 16.45 & 0.97\ $|-1\rangle$               + 0.25\ $|+1\rangle$\\
\hline 
\end{tabular}
\end{center}
\end{table}

The parameters given in Table~\ref{spintable} are determined by fitting the measured data at 35~K using the MF-RPA as implemented in the program $McPhase$\cite{rotter2004, rotter2002, rotter2012} within the data analysis environment provided by the program $Horace$\cite{ewings2016}.

In order to find starting parameters for the fit, we generate 10$^5$ random sets of exchange $J_{ij}$ and single-ion anisotropy $D_{\alpha}$ parameters and then use the previously reported 2~K data and linear spin-wave model\cite{toftpetersen2015} to filter out parameter sets which do not reproduce the low temperature dispersion. This procedure produces several sets of parameters which differ from the published set by having smaller $J_{ij}$ but larger $D_{\alpha}$ parameters, which although having slightly higher $\chi^2$ values still fit the 2~K data well. This is because the calculated spin-wave bandwidth and energy gap are governed by both the $J_{ij}$ and $D_{\alpha}$ parameters in a complex fashion, so that the same bandwidth and gap can result from smaller $J_{ij}$ and larger $D_{\alpha}$ parameters. However, as $H_{\mathrm{mf}}$ is determined by the exchange parameters $J_{ij}$, and has a larger effect on the energy $\Delta_{12}$ of the excited state transition, weaker $J_{ij}$ parameters fit the 35~K measurements better, as the $H_{\mathrm{mf}}$ from the parameters of Ref.~\onlinecite{toftpetersen2015} overestimates the energy of the excited state mode.

Furthermore, the MF-RPA predicts the physical properties better, since the calculated transition temperature in the mean-field approximation is $T_N^{\mathrm{mf}}=73$ K for the parameters of Ref.~\onlinecite{toftpetersen2015} compared to $T_N^{\mathrm{mf}}=62$ K for the set obtained from the fit of the neutron spectrum at 35~K, as presented in Figure~\ref{calculatedspinwave}(d-f). The calculated spectrum gives four excitation branches at 35~K rather than the three observed. The fourth mode, calculated to have an energy of 3.7 meV, corresponds to an excited state transition from the $n=2$ to the $n=4$ level, and is expected to have one fifth the intensity of the 4.5 meV excitation (from $n=1$ to $n=2$), with a calculated cross-section of 7 mb/sr/Fe$^{2+}$ compared to 36 mb/sr/Fe$^{2+}$. Thus it is probably too weak to be observed experimentally.

Inspection of the fitted exchange parameters show that the nearest-neighbour superexchange interaction $J_{bc}$ between Fe$^{2+}$ ions via a single oxygen ligand is more dominant than previously thought, with $J_{bc}/J_b\approx$5 (this study) rather than $J_{bc}/J_b\approx$2.5 (previous study\cite{toftpetersen2015}). Furthermore, $J_{b}$ and $J_{ab}$ have similar magnitudes and are both larger than $J_{c}$ and $J_{ac}$ which are extremely weak but non-zero. These next nearest neighbour interactions are mediated by superexchange via two oxygen ligands, but whilst $J_{c}$ and $J_{ac}$ couple Fe$^{2+}$ ions lying in octahedra which are tilted in opposite directions giving Fe-O-O-Fe bond angles closer to 90$^{\circ}$, the FeO$_6$ octahedra linked by $J_b$ and $J_{ab}$ are tilted in the same direction, resulting in more obtuse bond angles and hence greater overlap for the oxygen $p$-orbitals.


However, a potential shortfall of modeling the spin-orbit and crystal field interactions using the effective SIA parameters $D_a$ and $D_c$ is that it ignores higher order and odd-component crystal field terms which are permitted here because of the low symmetry of the crystallographic site occupied by the Fe$^{2+}$ ions. Using the full crystal field Hamiltonian, 
\begin{equation}
\label{duc2}
\mathcal{H}^i_{\mathrm{CF}} = \lambda \mathbf{L}_i \cdot \mathbf{S}_i  + \sum_{k=2,4} \sum_{q=-k}^k B_k^q O_k^q,
\end{equation}
\noindent in place of $\mathcal{H}_{\mathrm{SIA}}$ requires many more parameters, including  on-site spin-orbit coupling, $\zeta$, and the crystal field parameters $B_k^q$. We choose to restrict $\lambda$=12.75~meV to the free-ion value determined by optical spectroscopy and atomic calculations\cite{chakravarty1980, dunn1961}, and use the point-charge model to determine the $B_k^q$ parameters,  including charges within a cutoff range of 3.3 \AA~from the Fe$^{2+}$ ions. The point charge model has known shortcomings, such as not accounting for charge transfer or bonding effects, but allows us to reduce the number of parameters to three: the effective charges on each ligand atomic species. Starting parameters are obtained by requiring the point charges to approximately reproduce the energy level scheme in Table~\ref{zeemanlevels}. Fitting the 35~K inelastic dataset with this new point charge model, we obtained $q_{\mathrm{O}}$=-1.87(3) $|e|$ for the oxygen ligands, $q_{\mathrm{P}}$=0.58(3) $|e|$ for the phosphorous and $q_{\mathrm{Li}}$=0.04(2) $|e|$ for the lithium. Thus, it seems that the flowing Li  ions contribute very little to the anisotropy of the iron spins, as may be expected, whilst the largest effect is due to the distorted oxygen octahedron. The calculations does give the $b$ direction as the easy direction, but overestimates the ordered moment $\mu_{\mathrm{calc}}$=4.67 $\mu_B$ compared to a measured value of 4.09(4) $\mu_B$\cite{toftpetersen2015}.

More importantly, the calculated inelastic neutron spectrum is virtually identical to that shown in Figure~\ref{calculatedspinwave}(d-f) using the SIA parameters $D_a$ and $D_c$. Thus, the higher order crystal field terms appear to have little effect on the dispersion or intensities of the spin waves or excited state transition. This can be understood by reference to Table~\ref{zeemanlevels}, where we see that the ground state ($n$ = 0) and the first excited state ($n$ = 1) are both nearly pure states, that is the $n=0$ state is 99\% $|S_z = + 2\rangle$ and the $n=1$ state is 98\% $|S_z = + 1\rangle$, so the wavefunctions and hence the transition matrix elements which determine dispersion and scattering intensity are dominated by the ordered phase Zeeman field.

\begin{table}[htb]
\begin{center}
    \caption{\label{calculatedmoments} Calculated and reported moment directions and sizes for Li$M$PO$_4$. Calculations for Co\cite{tian2008}, Ni\cite{jensen2009}, and Mn\cite{toftpetersen2012}  are based on reported values of exchange parameters.  Note that for small amount of canting present in LiFePO$_4$ has been omitted here\cite{toftpetersen2015}.} 
\begin{tabular}{|c | c | c |}
\hline
$M$ & Calc. moment  & Exp. moment \\
 &  ($a,b,c$) (in $\mu_B$) &  ($a,b,c$) (in $\mu_B$) \\
\hline
Fe & (0, 4.67, 0) & (0, 4.09, 0)\cite{toftpetersen2015, rousse2003} \\
Co & (0, 3.87, 0) & (0, 3.35, 0) \cite{toftpetersenthesis} \\
 Ni & (0.41, 0, 2.80)& (0.3, 0, 2.2) \cite{jensen2009}\\
Mn & (5.00, 0, 0) &   (4.29,0,0)\cite{julien2006, toftpetersen2012}\\

\hline 

\end{tabular}\\

\end{center}
\end{table}

Nonetheless, the point charge model does show us that the $b$ easy-axis direction and anisotropic hard plane can be explained by the distorted geometry of the oxygen octahedron surrounding the Fe$^{2+}$ spins. Indeed, applying the same point charge model as fitted to the inelastic neutron scattering data from LiFePO$_4$ to other lithium-orthophosphates satisfactorily reproduces the measured ordered moment directions in Li$M$PO$_4$, although the magnitude of the ordered moments are overestimated by MF-RPA. Table \ref{calculatedmoments} provide a list of magnetic moment sizes and orientations calculated using the point charge model, with comparison to the reported values for Li$M$PO$_4$ ($M$=Fe, Co, Ni, and Mn).



\subsection{Spin-orbit splitting}
\begin{figure}
\includegraphics[width = 80mm]{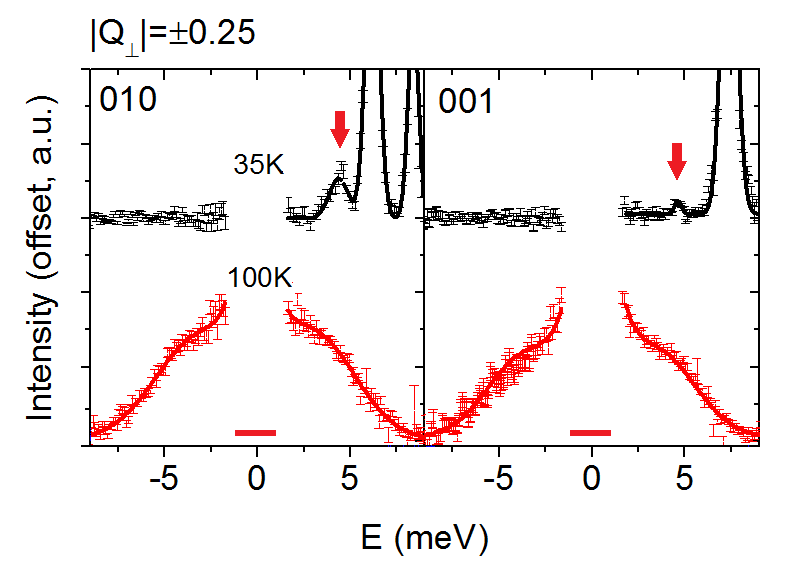}
\caption{\label{hightemperature1}Energy transfer at ($010$) and ($001$) for $T =$ 35 and 100~K. The integrated $Q$ range is $\pm 0.25$ along each direction. The red arrows indicates the extra excitation observed near 4 meV for 35~K along both directions. The thick red bar represents approximately the instrumental resolution width.}
\end{figure}

\begin{figure}
\includegraphics[width = 80mm]{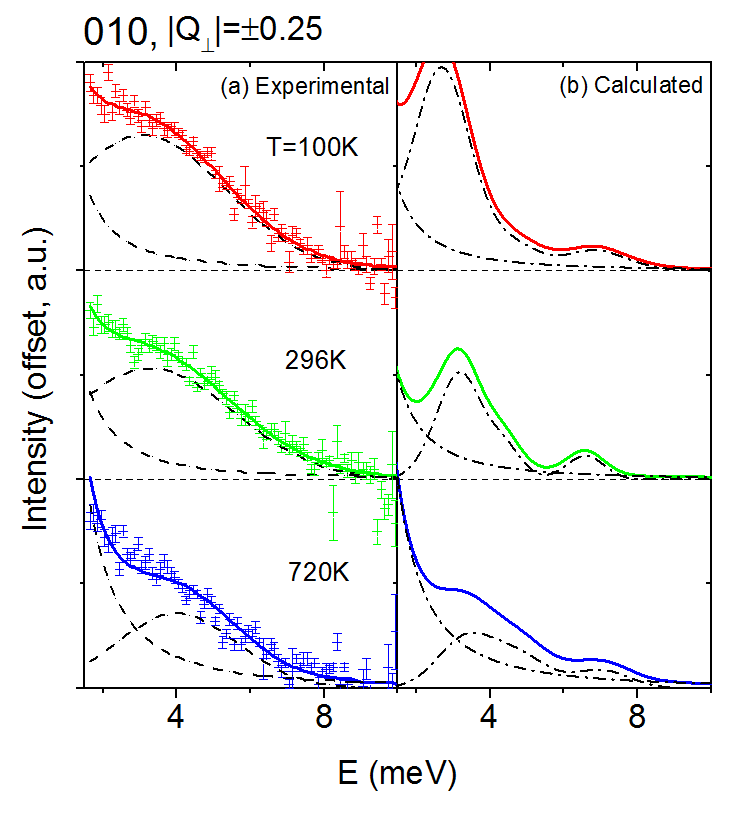}
\caption{\label{hightemperature2}(a): Measured energy transfer at ($010$) for $T$ = 100, 296 and 720~K. The dashed lines show the fitted Lorentzian centered at $E = 0$ and a Gaussian centered near 4 meV. The solid colored lines show the sum of the fitted Lorentzian and Gaussian curves. (b): Calculated energy transfer at ($010$) for the same temperatures.}
\end{figure}


Figure~\ref{hightemperature1} shows the energy responses of INS intensity around (010) and (001) at $T = 35$ and 100~K.  In order to correctly analyze the energy response, the negative energy transfer portion (i.e. neutron energy gain) of the data have been processed using the principle of detailed balance\cite{berk1993}.  The spectrum at $T = 35$ K, which is below $T_N$ and in the spin ordered phase, shows well defined peaks that correspond to the spin wave branches. The red arrows indicate the extra excitations observed near 4.5 meV as shown above. Above $T_N = 50K$ the system becomes paramagnetic and both the magnon dispersions and the hybrid 4.5 meV excitation disappear. 

Figure~\ref{hightemperature2}(a) shows a closer look of the INS energy responses at higher temperatures 100, 296, and 720~K. At these temperatures the spectra are relatively broad in energy, which includes a Lorentzian component centering at 0 meV and a Gaussian component centering near 4 meV, all represented as dashed lines in Figure~\ref{hightemperature2}(a). The Lorentzian response is expected of magnetic fluctuations that are quasi-elastic in nature and short-lived in time, however, there is little change in intensity and width over a temperature range many times $T_N$.


Broad excitations are detected around 3.4(4) meV at 100~K and 4.2(7) meV at 720~K, determined by the fitted Gaussian curves as represented by the dashed Gaussian lines in Figure~\ref{hightemperature2}(a). Since the internal fields generated by the ordered moments should be absent at these temperatures, these excitations cannot be the same as that discussed above, seen at 35~K. However, as shown in Table~\ref{zeemanlevels}, in the paramagnetic phase the spin-orbit and crystal field interactions combined still splits the $S=2$ levels, for instance the energy for a $n$ = 2 to 0 transition is 3.39 meV, which may be what is observed at 100~K. At 720~K, we should expect that all the excited states are populated, allowing higher transitions to have greater spectral weight, and shifting the observed peak to a higher energy. 
However, the data is much broader than the instrumental resolution, and MF-RPA theory we have used to analyse the low temperature data does not account for thermal broadening of the excitations. Instead we have convoluted the calculation with Gaussian curves with a full width at half maximum of 1.2~meV to obtain the curves in figure~\ref{hightemperature2}, whereas instrumental resolution at these energy transfers is expected to be less than 0.5~meV. It can be seen that this still does not fully account for the measurements. In addition, the calculations predict that the intensity should fall off much faster with increasing temperature than is observed: the calculated intensity ratio of the peak area at 720~K vs 100~K is 0.35, which is half that observed ($I_{\mathrm{720K}}/I_{\mathrm{100K}}\approx0.7$). Thus, whilst the energies of the peaks observed at high temperatures may be satisfactorily explained by the MF-RPA theory, their intensity and linewidths require a more sophisticated approach.

\section{Summary}
In summary, we present inelastic neutron scattering results in LiFePO$_4$  that complements a previous spin wave excitations study. In particular, we find an extra excitation at 4.5 meV at $T = 35$ K $< T_N$ that is nearly dispersionless and is most intense around magnetic zone centers. We show that these excitations correspond to transitions between thermally occupied excited states of Fe$^{2+}$ due to  splitting of the $S=2$ levels that  arise from the effects of the crystal field and spin-orbit interactions amplified by the highly distorted nature of the oxygen octahedron surrounding the iron spins. Above $T_N$ the magnetic fluctuations are observed to relatively high temperatures, with little temperature dependence between 100 and 720~K. Additional excitations, broad in energy, are observed around 4 meV that are due to the single-ion splittings caused by the spin-orbit and crystal field interactions. These excitations weaken slightly at 720~K compared to 100~K, consistent with the calculated cross-sections from our single-ion model. Our theoretical analysis using the MF-RPA model provides detailed spectra of the $d-$shell in LiFePO$_4$ and also enables estimates of the average ordered magnetic moment and $T_N$. Applying it to spin-wave results of other members of the Li$M$PO$_4$ ($M=$ Mn, Co, and Ni)  compounds provides reasonable ordered moments and transition temperatures showing the approach is robust.

\section{Acknowledgement}

We thank Jens Jensen for very detailed and fruitful discussions.
Research at Ames Laboratory is supported by the U.S. Department of Energy, Office of Basic Energy Sciences, Division of Materials Sciences and Engineering under Contract No. DE-AC02-07CH11358. Use of the Spallation Neutron Source at the Oak Ridge National Laboratory is supported by the U.S. Department of Energy, Office of Basic Energy Sciences, Scientific Users Facilities Division.

\bibliography{paper}
\end{document}